\documentclass[12pt,
	aps,
	letterpaper,
	preprintnumbers,
	amsmath,
	amssymb,
	nofootinbib,
	showkeys,
	notitlepage
]{revtex4-1}

\usepackage{graphicx}   
\usepackage{longtable}
\usepackage{xcolor}
\usepackage{lineno,hyperref}
\usepackage[normalem]{ulem}

\usepackage{graphicx}   

\begin{document}
\title{Inter-relations between additive shape invariant superpotentials}

\author{Jeffry V. Mallow}
\email{jmallow@luc.edu}
\author{Asim Gangopadhyaya}
\email{agangop@luc.edu}
\author{Jonathan Bougie}
\email{jbougie@luc.edu}
\author{Constantin Rasinariu}
\email{crasinariu@luc.edu}
\affiliation {Department of Physics, Loyola University Chicago, Chicago, IL 60660, U.S.A}

\date{\today}

	
\begin{abstract}
All known additive shape invariant superpotentials in nonrelativistic quantum mechanics belong to one of  two categories:  superpotentials that do not explicitly depend on $\hbar$, and their $\hbar$-dependent extensions. The former group themselves into two disjoint classes, depending on whether the corresponding Schr\"odinger equation can be reduced to a hypergeometric equation (type-I) or a confluent hypergeometric equation (type-II). All the superpotentials within each class are connected via point canonical transformations. Previous work \cite{Gangopadhyaya} showed that type-I superpotentials produce type-II via limiting procedures. In this paper we develop a method to generate a type I superpotential from type II, thus providing a pathway to interconnect all known additive shape invariant superpotentials.
\end{abstract}
	
\maketitle
		
\section{Introduction and Background}

\subsection{Supersymmetric Quantum Mechanics}
Supersymmetric quantum mechanics (SUSYQM)  is a  generalization of the Dirac-F\"ock ladder method for the harmonic oscillator \cite{Witten,Solomonson,CooperFreedman}.  In SUSYQM, a general hamiltonian $H_-$ is written in terms of ladder-operators ${\cal A}^+ \equiv -\frac{\hbar}{\sqrt{2m}}  \frac{d}{dx}+ W(x,a)$ and ${\cal A}^- \equiv \frac{\hbar}{\sqrt{2m}}  \frac{d}{dx}+ W(x,a)$, where the function $ W(x,a)$, a real function of $x$ and a parameter $a$, is known as the superpotential. Henceforth, we set $2m = 1$.  The hamiltonian $H_-$ is given by 
\begin{eqnarray}
H_-= {\cal A}^+\, {\cal A}^-
&=& \left(  -\hbar \frac{d}{dx}+ W(x,a) \right)  ~\left( \hbar \frac{d}{dx}+ W(x,a)\right) \nonumber \\
&=&  - \hbar^2 \frac{d^2}{dx^2}+W^2(x,a)  -  \hbar\,\frac{dW(x,a)}{dx}~ \nonumber \\
&=& 
- \hbar^2 \frac{d^2}{dx^2}+V_-(x,a)
\label{A+A-},
\end{eqnarray}
where $V_-(x,a) = W^2(x,a)  - \hbar~ {d W}/{dx}$.  
The product of operators ${\cal A}^-  {\cal A}^+$ generates another hamiltonian  $H_+=  -\hbar^2 \frac{d^2}{dx^2}+V_+(x,a)$ 
with $V_+(x,a) = W^2(x,a)  +  \hbar ~{d\, W}/{dx}$. These two hamiltonians are related by ${\cal A}^+H_+ = H_-\,{\cal A}^+$ and 
${\cal A}^-H_- = H_+\,{\cal A}^-$, which leads to the following isospectrality relationships among their eigenvalues and eigenfunctions for all integer $n \geq 0 $:
\begin{eqnarray}
E^{-}_{n+1} =E^{+}_{n}; \qquad
\frac{~~~{\cal A}^- }{\sqrt{E^{+}_{n} }} ~\psi^{(-)}_{n+1} = ~\psi^{(+)}_{n} \,,~ ~ {\rm and} \quad
\frac{~~~{\cal A}^+}{\sqrt{E^{+}_{n} }}~\psi^{(+)}_{n} = ~  \psi^{(-)}_{n+1}. \label{isospectrality}
\end{eqnarray}
Thus, if we knew the eigenvalues and eigenstates of the hamiltonian $H_-$,\footnote{Since $H_-$ is a semi-positive-definite operator, its ground state energy $E_0$ is either zero or positive. When $E_0=0$, supersymmetry is said to be unbroken.} we would automatically know the same for the hamiltonian $H_+$, and vice-versa. For unbroken SUSY, if a superpotential $W(x,a_i)$ obeys a particular constraint known as ``shape invariance'', then the eigenvalues and eigenfunctions for both hamiltonians can be determined separately. In this manuscript, we will consider only unbroken SUSY.
\subsection{Shape Invariance}
Let us consider a set of parameters $a_{i}$, $i=0,1, 2,\cdots$, with $a_{0}=a$, and $a_{i+1}=f(a_{i})$, where $f$ is a function of $a_i$.  A superpotential $W(x,a_i)$ is shape 
invariant if it obeys the following condition \cite{Infeld,Miller,Gendenshtein1,Gendenshtein2}:
\begin{eqnarray}
W^2(x,a_i)  +  \hbar \,\frac{d\, W(x,a_i)}{dx}+g(a_i) = 
W^2(x,a_{i+1})  -  \hbar\, \frac{d\, W(x,a_{i+1})}{dx}+g(a_{i+1}).~ \label{SIC1}
\end{eqnarray}
The eigenvalues and eigenfunctions are given by \cite{Cooper-Khare-Sukhatme,Gangopadhyaya-Mallow-Rasinariu}
\begin{eqnarray}
E_n^{(-)}(a_0)=g(a_n)-g(a_0)~~{\rm
	for}~n\geq0~,\nonumber
\label{energy_sic} 
\end{eqnarray}
and
\begin{eqnarray}
&\psi^{(-)}_{n}(x,a_0)\sim{\cal A}^+{(a_0)} ~ {\cal A}^+{(a_1)}  \cdots  {\cal A}^+{(a_{n-1})} ~ \psi^{(-)}_0(x,a_n)~,\nonumber
\end{eqnarray}
where $\psi^{(-)}_0(x,a_n) \sim \exp\left\{-\frac1\hbar\, \int^x W(y,a_n)\,dy\right\}$.
This solvability of all additive shape invariant systems, which stems from Eq. (\ref{SIC1}), can be related to underlying potential algebras of the systems \cite{Gangopadhyaya_algebra1,Balantekin1,Gangopadhyaya_algebra2,Gangopadhyaya_algebra3,Balantekin2}.

Hereafter, we consider the case of translational or additive shape invariance: $a_{i+1}=a_{i}+\hbar$.

\section{Shape Invariant Superpotentials}
Shape invariant systems are of great importance in quantum mechanics due to their exact solvability; hence, it is desirable to determine as many shape invariant superpotentials (SISs) as possible. All SISs obey Eq. (\ref{SIC1}), which is a non-linear difference-differential equation. Several investigators have found solutions to this equation \cite{Infeld,Gendenshtein1,Dutt,CGK}. The authors of Ref. \cite{Gangopadhyaya_NPDE,Bougie2010,symmetry} reduced Eq. (\ref{SIC1}) to two local partial differential equations (PDEs) and proved that the list of SISs listed in \cite{Infeld,Gendenshtein1,Dutt,CGK} is complete, under the assumption that $W(x,a)$ does not depend explicitly on $\hbar$. The set of superpotentials generated by solving the two PDEs was called   ``conventional''.

In 2008, two additional shape invariant superpotentials were discovered  \cite{Quesne1,Quesne2} that were not included in previous lists of conventional superpotentials. These superpotentials were then generalized in Ref.\cite{Odake1, Odake2, Tanaka, Odake3, Odake4, Quesne2012a, Quesne2012b}, and some of their properties have been further studied \cite{Ranjani1,Ramos2011}. Since these superpotentials could not be generated from the two PDEs, they must contain explicit $\hbar$-dependence.  In Ref. \cite{Bougie2010,symmetry}, the authors showed that these superpotentials obey an infinite set of PDEs.  In this section, we will describe how to generate these shape invariant systems from the PDEs.

\subsection{Conventional Superpotentials}
We begin with conventional superpotentials, for which $W(x,a_i)$ has no explicit dependence on $\hbar$; i.e., any dependence on $\hbar$ enters only through the linear combination $a_{i+1}=a_i+ \hbar$. Since Eq. (\ref{SIC1}) must hold for an arbitrary value of $\hbar$, we can expand the equation in powers of $\hbar$, and require that the coefficient of each power vanishes, leading to the following two independent equations \cite{Bougie2010,symmetry} :
\begin{eqnarray}\vspace{6pt} W \, \frac{\partial W}{\partial a} - \frac{\partial W}{\partial x} + \frac12 \, \frac{d g(a)}{d a} = 0~
\label{PDE1}
\end{eqnarray} and
\begin{eqnarray}
\frac{\partial^{3}}{\partial a^{2}\partial x} ~W(x,a)= 0\label{PDE3}~.
\end{eqnarray}
The general solution to Eq. (\ref{PDE3}) is
\begin{eqnarray}
W(x,a)=a\cdot \chi_1(x)+\chi_2(x)+u(a)~\label{GeneralSolution}.
\end{eqnarray}
When combined with  Eq. (\ref{PDE1}), this solution reproduces  the complete family of  conventional superpotentials, as shown in Table (\ref{table:conventional})  \cite{Bougie2010,symmetry}.

These conventional superpotentials generate special cases of the Natanzon potentials \cite{Natanzon,Cooper-Khare-Sukhatme,Gangopadhyaya-Mallow-Rasinariu}. They fall into one of two categories, depending on whether the corresponding Schr\"odinger equation can be reduced to a hypergeometric equation (type-I) or a confluent hypergeometric equation (type-II) \cite{DeDuttSukhatme,Gangopadhyaya}.

\begin{table} [htbp]
	\begin{center}
		\begin{tabular}{||c|l|l|c||}
			\hline
		~~~~~~~&~Name                  & ~ Superpotential                      & ~Type~ \\ \hline
		1	&~Scarf (Hyperbolic)    & ~  $A\tanh x + B\, {\rm sech}\,x$       &   I    \\ \hline
		2	&~Gen. P\"oschl-Teller  & ~  $A\coth r- B\, {\rm cosech}\,r $ ~   &   I    \\ \hline
		3	&~Scarf (Trigonometric) & ~  $ A\tan x - B\, {\rm sec}\,x$        &   I    \\ \hline
		4	&~Rosen-Morse I         & ~ $-A\cot x - \frac{B}{A}$            &   I    \\ \hline		
		5	&~Rosen-Morse II        & ~ $A\tanh x + \frac{B}{A}$            &   I    \\ \hline
		6	&~Eckart                & ~$-A\coth r + \frac{B}{A}$            &   I    \\ \hline		
		a	&~Morse                 & ~ $A- B\, e^{-x}$                         &   II   \\ \hline
		b	&~3-D oscillator        & ~ $\frac12 \omega r - \frac{\ell}{r}$ &   II   \\ \hline
		c	&~Coulomb               & ~ $\frac{e^2}{2\ell}- \frac{\ell}{r}$ &   II   \\ \hline
		d	&~Harmonic Oscillator~  & ~ $\frac12 \omega x$                  &   II   \\ \hline
		\end{tabular}
		\caption{The complete family of $\hbar$-independent additive shape-invariant superpotentials.} 
		\label{table:conventional}
	\end{center}
\end{table}

\subsection{Extended Superpotentials}
Table (\ref{table:conventional}) lists the exhaustive set of superpotentials that are $\hbar$-independent.  However, this list does not include the $\hbar$-dependent shape invariant superpotentials reported in \cite{Quesne1,Quesne2}, or their generalizations  \cite{Odake1, Odake2, Tanaka, Odake3, Odake4}. All known $\hbar$-dependent superpotentials can be written as $W(x,a,\hbar) = W_0(x,a)+W_h(a,x,\hbar)$, where the kernel $W_0$ is one of the conventional superpotentials listed in Table \ref{table:conventional}, and $W_h$ is an explicitly $\hbar$-dependent extension of that kernel. For example, one superpotential found in \cite{Quesne1} can be written as
\begin{equation}
\label{eq:Quesne-ext}
W(r, \ell, \hbar)= \frac{1}{2}\, \omega r-\frac{\ell}{r} +\left(\frac{2 \omega r \hbar}{\omega r^{2}+2 \ell-\hbar}-\frac{2 \omega r \hbar}{\omega r^{2}+2 \ell +\hbar}\right) ~,
\end{equation}
where $W_0 = \frac 12 \, \omega r - \frac{\ell}{r}$ is the conventional superpotential of the 3-D oscillator, and the term in parenthesis is $W_h$, the $\hbar$-dependent extension.

Because extended superpotentials depend explicitly on $\hbar$, we can expand them in powers of $\hbar$:
\begin{eqnarray}
\vspace{6pt}
W(x, a, \hbar) = \sum_{j=0}^\infty \hbar^j W_j(x,a).~ \label{W-hbar}
\end{eqnarray}
Since  $a_0=a$ and $a_1=a+\hbar$,  we also have
\begin{eqnarray}
W\left(x,a_1,\hbar\right)=\sum_{j=0}^\infty \sum_{k=0}^{j}\frac{\hbar^j}{k!}\frac{\partial^k W_{j-k}}{\partial a^k},\nonumber
\end{eqnarray} 
which we then substitute back into Eq. (\ref{SIC1}).  Since Eq. (\ref{SIC1}) must hold for any value of $\hbar$, we set the coefficients of the series for each power of $\hbar$ equal to zero. 
This gives, for $j=1$
\begin{eqnarray}
2\frac{\partial W_{0}}{\partial x}-\frac{\partial }{\partial a} \left(W_{0}^2+g  \right) = 0,
\label{lowestorder}
\end{eqnarray}
and for $j \geq 2$
\begin{equation}
2\,\frac{\partial W_{j-1}}{\partial x}
 -\sum_{s=1}^{j-1} \sum_{k=0}^s \frac{1}{(j-s)!}
\frac{\partial^{j-s}}{\partial a^{j-s}} W_k\, W_{s-k} 
+  
\sum_{k=2}^{j-1} \frac{1}{(k-1)!}
\frac{\partial^{k}\, W_{j-k}}{\partial a^{k-1} \,\partial x}=0 
 ~.
\label{higherorders}
\end{equation}
In Refs. \cite{Bougie2010} and \cite{symmetry} the authors explicitly generated the extended superpotential (\ref{eq:Quesne-ext}) from these partial differential equations.

\section{Known Relationships Between Shape-Invariant Superpotentials}

In this section we discuss the known connections between SISs. They are: point canonical transformations, projections, and isospectral extensions.

\subsection{Point Canonical Transformations}

We begin with the relationships between the various conventional superpotentials that are connected via point canonical transformations (PCTs). A PCT comprises a change of the independent variable $x$ and an associated multiplicative transformation of the wavefunction in a Schr\"odinger equation, such that it generates a new Schr\"odinger equation \cite{Bhattacharjie,DeDuttSukhatme}. 

For a change of variable  from $x \to z$, where $x=u(z)$ and a corresponding change in wave function that relates the new wave function $\tilde{\psi} $ to the old by  $\psi(x) = \tilde{\psi}(z)\,\sqrt{du/dz} $,  the Schr\"odinger equation 
\begin{equation}
	-\frac{d^2 \psi(x)}{dx^2} + V(x,a_i)\, \psi(x)  = 
	E(a_i) \psi(x)
	\label{eq.1}
\end{equation}
transforms into: 
\begin{equation}
	\left[ 	-\frac{d^2 }{dz^2}
	+ \left\{ 
	\dot{u}^{\,2}\, \left[ V(u(z),a_i) - E(a_i) \right] + 
	{\frac 12}
	\left( \frac{3\ddot{u}^{\,2}}{ 2\dot{u}^{\,2}} - \frac{\dddot{u}}{\dot{u}}  \right)
	\right\} 
	\right] \tilde{\psi} \left( z \right) = 0.
	\label{eq:second_order}
\end{equation}\\
where $\dot{u}=du/dz$, etc. For Eq. (\ref{eq:second_order}) to be a Schr\"odinger equation, an energy term must emerge from the expression 
$ \dot{u}^{\,2} \, \left[ V(u(z),a_i) - E(a_i)\right]$; i.e., it must have a term that is independent of $z$.   This condition constrains the choices for the function $u(z)$.  

The six type-I superpotentials are characterized by corresponding Schr\"odinger equations that can be transformed into a hypergeometric equation. If we consider the one-dimensional harmonic oscillator to be a simplified case of the 3-D oscillator ($\ell=0$) \footnote{Note that setting $\ell=0$ removes the singularity at the origin and hence enlarges the domain to the entire real axis.}, we have three type-II superpotentials, which correspond to the confluent hypergeometric equation. Each type-I superpotential can be mapped to each other type-I superpotential via PCTs, and each type-II superpotential can be similarly mapped to each other type-II superpotential \cite{CooperGinocchioWipf, DeDuttSukhatme, Levai}. The corresponding PCTs, illustrated in Figure \ref{fig:Hexagon}, are given by 
$
T_{12}\!:\! \{x\to r+ i \pi /2 \},\, 
T_{23}\!:\! \{ r\to i x+ i \pi /2\},\, 
T_{34}\!:\! \{x\to \cos ^{-1}(\text{cosec } x) \},\, 
T_{45}\!:\! \{ x\to \pi /2 +i x \}, \,
T_{56}\!:\! \{ x\to -r+ i \pi/2\},\, 
T_{61}\!:\! \{ r\to \coth ^{-1}(i \sinh x) \},\, 
T_{ab}\!:\! \{x\to -2 \ln r\},\, 
T_{bc}\!:\! \{ r\to \sqrt z\},\, 
$
and
$
T_{ca}\!:\! \{ r\to \exp (-x)\}\,.
$

\subsection{Projections}

PCTs cannot transform type-I to type-II or vice-versa. 
The hypergeometric differential equation, corresponding to type-I superpotentials, has three regular singular points.  With suitable limits, two of the singularities merge, and the equation reduces to a confluent hypergeometric equation, connected to the type-II superpotential. Thus, these limiting procedures generate ``projections'' from type-I to type-II superpotentials, making it possible to move from one type to another, albeit in only one direction, as shown in 
Table~\ref{table:reductions}, and in Figure \ref{fig:Hexagon}.

\begin{table}[htbp]
	\vspace{0cm}
		\begin{center}
\footnotesize
\begin{tabular}{||l|l|l||}  \hline
	{\bf Type-I Superpotential}   & {\bf Projection }                     & {\bf Type-II Superpotential } \\ \hline
	{\bf 1) Scarf (Hyperbolic)}   & $P_{1a}:$	                          & {\bf a) Morse } 	          \\
	$W(x)=A\tanh\left( x + \beta\right)+B~\textrm{sech}\left( x + \beta\right)$       &                       
	$A\to A$                                                                  & 
	$W(x)=A- B\, e^{-x}$                                                                                       \\
	$-\infty<x<\infty, \;A>0 $ 	                                                      & 
	$B\to - B\, {e^\beta \over 2}$                                              & 
	$-\infty<x<\infty $                                                                                   \\
	$E_n=A^2-\left(A-n \hbar \right)^2$  &	$\beta\to \infty$             & $E_n=A^2-\left(A-n \hbar \right)^2$  \\ \hline
	{\bf 2) Generalized P\"oschl-Teller }  & $P_{2a}$:                    & {\bf a) Morse }               \\
	$W(r)=A\coth\left(\alpha r + \beta\right)-B~\textrm{cosech}\,\left(\alpha r + \beta\right)$         &
	$A \to A $ 	                                                              & 
	$W(x)=A-B\,e^{ -x}$                                                                                      \\
	{~~~}        & $B\to B\, {e^\beta \over 2}$, $\alpha \to 1$                   & $-\infty<x<\infty $           \\ 
	{~~~}        & $\beta\to \infty$, $r \to x$           & $E_n=A^2-\left(A-n\hbar \right)^2$  \\ \cline{2-3}
	{~~~}                         & $P_{2b}:$                             & {\bf b) 3-D Oscillator }      \\
	$-\frac{\beta}{\alpha} < r<\infty$			                                                      & 
	$A \to \left({\omega \over \alpha} - \frac{\alpha\ell}{2}\right)$         & 
	$W(r)=\frac{1}{2}\omega r -\frac{\ell}{r}$                                                            \\
	$E_n=A^2-\left(A-n\alpha\hbar \right)^2$					                                  &
	$B \to \left({\omega \over \alpha} +  \frac{\alpha\ell}{2}\right)$        & 
	$0<r<\infty$                                                                                          \\
	$A<B$       & $\beta\to 0$, $\alpha \to 0$            & $E_n=2n\omega\hbar $                \\ \hline
	{\bf 3)  Scarf (Trigonometric) }     & $P_{3b}:$                      & {\bf b) 3-D Oscillator}       \\
	$W(x)=A \tan\left(\alpha x\right) - B\sec\left(\alpha x\right)$		              &
	$A \to \left({\omega \over \alpha} + \frac{\alpha\ell}{2}\right)$         & 
	$W(r)=\frac{1}{2}\omega r -\frac{\ell}{r}  $                                                          \\
	$-{\pi \over 2\alpha}<x<{\pi \over 2\alpha}, 	~A>B$                             &
	$B \to \left({\omega \over \alpha} -  \frac{\alpha\ell}{2}\right)$        & 
	$0<r<\infty$                                                                                          \\
	$E_n=\left(A+n\alpha\hbar\right)^2-A^2$                                                & 
	$x\to r+{\pi \over 2\alpha}$, $\alpha \to 0\; $			          & 
	$E_n=2n\omega\hbar $                                                                                       \\ \hline
	{\bf 4) $  $Rosen-Morse I}    & $P_{4c}:$		                      &	{\bf c) Coulomb}		      \\
	$W(x)=-A\cot\left(\alpha x\right)-\frac{B}{A}$	                                  &
	$A\to \alpha \ell$					                                      & 
	$W(r)=\frac{e^2}{2\ell}	-\frac{\ell}{r}$	                                                          \\
	$ 0 < x <\frac{\pi}{\alpha} $	                              &
	$ B \to - {\alpha \over 2} e^2$		 				                      & 
	$0<r<\infty $ 	                                                                                      \\
	$E_n=- A^2+\left(A+n\alpha\hbar\right)^2+{B^2 \over A^2} -{B^2 \over {(A+n\alpha\hbar)}^2}$ &
	$\alpha \to 0$, $x\to r$	                                          & 
	$E_n={e^4 \over {4\hbar^2}}\left( {1\over l^2}-{1\over (n+l)^2}\right)$	                              \\ \hline
	{\bf 5) $  $Rosen-Morse II}      & {~~} ---                           & {~~} ---	                  \\
	$W(x)= A \tanh(x) + \frac BA$    & {~~}                               & {~~}                          \\
	$-\infty<x<\infty $, $B < A^2$   & {~~}                               & {~~}                          \\
	$E_n= A^2-\left(A-n\hbar\right)^2-\frac{B^2}{(A-n\hbar)^2} +\frac{B^2}{A^2}$ & {~~} &                 \\ \hline
	{\bf 6)  Eckart}              & $P_{6c}:$ 		                      & {\bf c) Coulomb}              \\
	$W(r)=-A\rm{coth}\left(\alpha r\right)+\frac{B}{A}	$	                          &
	$A\to \alpha \ell$						                                  & 
	$W(r)=\frac{e^2}{2\ell}-\frac{\ell}{r} $                                                              \\
	$0<r<\infty,\; B>A^2,\; A>0$ & $ B \to {\alpha \over 2} e^2$  & $0<r<\infty $                 \\
	$E_n= A^2-\left(A+n\alpha\hbar\right)^2+{B^2 \over A^2} -{B^2 \over {(A+n\alpha\hbar)}^2}$  &
	$\alpha\to 0$                                                             & 
	$E_n={e^4 \over {4\hbar^2}}\left( {1\over l^2}-{1\over (n+l)^2}\right)$                                  \\ \hline
\end{tabular}
	\caption{Limiting procedures and redefinition of parameters relating type-I to type-II superpotentials. For projections, in each cell the order of operators should be carried out from top to bottom.}
	\label{table:reductions}
\end{center}
\end{table}

\subsection{Isospectral Extensions}
For extended superpotentials, the energy spectrum is given entirely by the $\hbar-$independent kernel $W_0$ \cite{symmetry}, and every known $\hbar-$dependent SIS contains a conventional SIS as its kernel.  These extended superpotentials can therefore be obtained from conventional superpotentials through an isospectral process that adds an $\hbar$-dependent term to the conventional superpotential while maintaining shape-invariance. In the limit $\hbar \to 0$, each reduces to its corresponding conventional counterpart.

\section{Generating a Pathway from Type-II to Type-I Superpotentials }

\begin{figure}[htpb]
	\centering
	\includegraphics[width=0.95\linewidth]{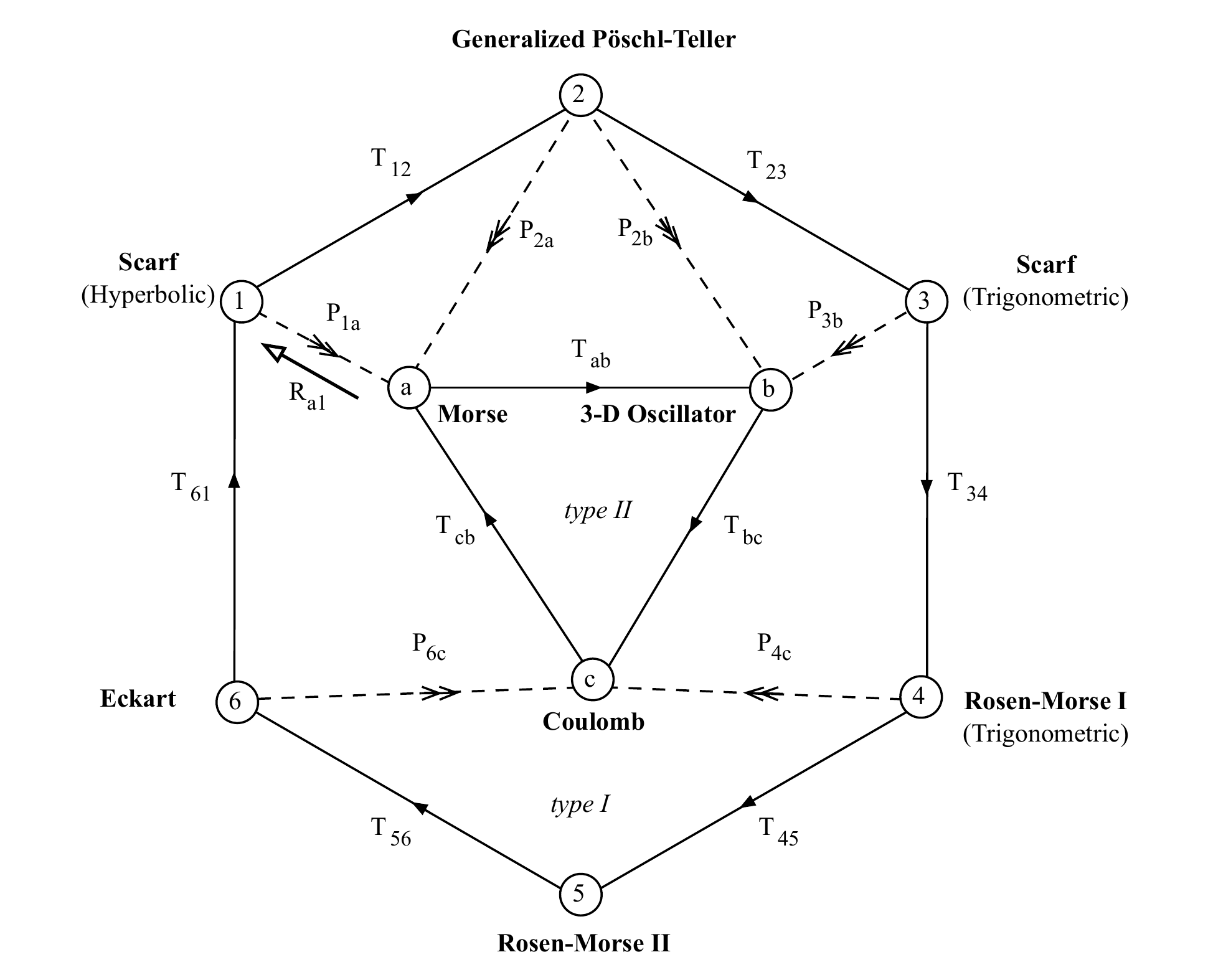}
	\caption{Inter-relations among conventional superpotentials. PCTs are represented by plain lines with arrows while the projections from type-I to type-II are represented by dashed lines with double arrows; the projection corresponding to each label is given in Table~\ref{table:reductions}. $R_{a1}$ will be discussed below.}
	\label{fig:Hexagon}
\end{figure}

We have seen that the six SISs of type-I are interconnected via PCTs; so are the three type-II SISs. Furthermore, type-I SISs reduce to type-II via projections. Graphically, these interrelations are illustrated in Figure \ref{fig:Hexagon}.  
%
Additionally, the extended SISs are obtained from conventional SISs by isospectral extension and they reduce to the conventional ones when $\hbar \to 0$.

However, we have no connection yet that will take us from a type-II to a type-I superpotential. So far, the connections between the SISs have been of three types: (i) PCTs, (ii) projections, and (iii) isospectral extensions. We now ask whether we can employ any of these three mechanisms to go from type-II to type-I. 

Of these three mechanisms, PCTs map superpotentials within a given type (I or II), but do not move between types. Projections  reduce the hypergeometric equation to the confluent hypergeometric equation, but do not do the reverse. This leaves the isospectral extension, which allows for the addition of terms to an initial kernel $W_0$. We choose this kernel to be Morse, because it is the only type-II SIS that is isospectral with type-I SISs. Therefore, if such a reverse path, denoted $R_{a1}$ in Figure \ref{fig:Hexagon},  exists, then it should start from Morse.

We proceed to construct an extension using  Eq. (\ref{higherorders}).
In Ref. \cite{Quesne2012b}, the author generated a quasi-exactly solvable extension of Morse and showed that it was not shape invariant. The strength of the isospectral extension method is that we can employ Eq. (\ref{higherorders}) term-by-term in order to generate a manifestly
shape-invariant solution. The Morse superpotential is
$$
W_0=-a - e^{-x}~,
$$
where, without loss of generality, we set $a\equiv-A<0$, and $B=1$\footnote{Note that $B \ne 1$ amounts to a simple translation in $x$.}.
Choosing $W_1=0$, the equation for $W_2$ reads
\begin{equation}
	\frac{\partial W_{2}}{\partial x}
	- \frac{\partial W_{0} W_2}{\partial a}  = 0~.\nonumber 
	\label{Eq_order3}
\end{equation}
A solution is
\begin{eqnarray}W_{2}(x,a) =e^{- x} \left( 2 P + Q e^{-2 x} 
	+ 2  a~ Q e^{- x}\right)~,
	\nonumber\end{eqnarray}
where $P$ and $Q$ are constant parameters.
Choosing $W_3$ to be zero, the equation for $W_4$ is 
\begin{equation*}
\begin{aligned}
	2\frac{\partial W_4}{\partial x}
	- 
	2\frac{\partial \left( W_0 W_4+\frac12 W_2^2\right) }{\partial a}  - 
	\frac{\partial }{\partial a}
	\left( \frac{\partial W_{2}}{\partial a} 
	\frac{\partial W_{0}}{\partial a}  \right) \\ 
	- 
	\frac13 \left(
	W_2 \frac{\partial^3 W_{0}}{\partial a^3} 
	+
	W_0 \frac{\partial^3 W_{2}}{\partial a^3} \right) 
	+ \frac{1}{2}\frac{\partial^3 W_{2}}{\partial x \partial a^2}
	= 0
\end{aligned}
\end{equation*}
The above equation is solved by  
\begin{eqnarray}
	W_{4}(x,a) = - Q e^{-3 x} \left( 2 P + Q e^{-2 x} 
	+ 2 a ~ Q e^{- x}\right)~.
	\nonumber\end{eqnarray}
Generalizing this process yields  
$W_{2k-1}=0$ and 
\begin{equation}
	W_{2k}= 
	\left(- Q\right)^{k-1} e^{-(2k-1) x} \left( 2 P + Q e^{-2 x} 
	+ 2 a ~  Q e^{- x}\right)\nonumber
\end{equation}
for all positive integers $k.$ 
Computing the infinite sum $\sum_{j=0}^\infty \hbar^j 
W_j(x,a)$, 
we obtain 
\begin{equation}
	W(x,a,\hbar) = -a -e^{- x}  +
	\frac{\hbar^2 \left( 2 Pe^{x} + 2 a\, Q + Q\, e^{- x} 
		\right)}{e^{2 x}+Q\,\hbar^2}~.
	\label{newpotential}
	\end{equation}

\noindent 
The shape invariance of this superpotential can be directly
checked. Substituting the above expression into Eq. (\ref{SIC1}) yields
\begin{equation}
	W^2(x,a) - W^2(x,a+\hbar)  +
	\hbar \frac{d\,}{dx} \left( W(x,a)+W(x,a+\hbar)\right) 
	=
	-\hbar (2 a + \hbar),
\end{equation}
which can be brought into the form of Eq. (\ref{SIC1}) by choosing $g(a) =
-a^2$.  This leads to the energy eigenvalues $E^{(-)}_n =
g(a+n\hbar)-g(a) = a^2 - (a+n\hbar)^2$. As expected, these values are the same as those of the Morse potential. Note that as $\hbar \to 0$, we recover the starting kernel, which is the Morse superpotential. 

The superpotential Eq. (\ref{newpotential}) was initially reported in Ref.\cite{Bougie2015} as a new $\hbar$-dependent extension of the Morse superpotential. However, here we show that it is in fact equivalent to the conventional Scarf hyperbolic superpotential. To do so, we absorb $\hbar$ in Eq. (\ref{newpotential}) into another set of parameters via the following transformations:
\begin{equation}
	\label{toScarf}
	\hbar^2 P \to P^{\,\prime} ,~
	\hbar^2 Q \to e^{2\beta}, ~
	(2 P^{\,\prime} -1)\,e^{-\beta} \to 2 B,~
	-a \to A,~
	x-\beta \to x ~.
\end{equation}
These transformations effectively map the ``extended'' superpotential (\ref{newpotential}) into the Scarf hyperbolic, a conventional type-I superpotential \, \footnote{A similar reduction could be obtained using the formalism in \cite{Ramos2000}  via a suitable choice of parameters.}:
\begin{equation}
	-a - e^{- x}  +
	\frac{\hbar^2 \left( 2 P\,e^{x} + 2a\, Q + Q\, e^{- x} 
		\right)}{e^{2 x}+Q\,\hbar^2}~
	\to W_{Scar\!f} = A \tanh x + B\,\textrm{sech}\, x~.
\end{equation}

Thus, in this case, rather than producing a new superpotential, this technique created a ``restricted extension'' $R_{a1}$, from a type-II to a type-I superpotential.  $R_{a1}$ was the missing link in the quest to provide a connection between all known additive shape invariant superpotentials. Now, we have a bidirectional way to connect any pair of known additive shape-invariant superpotentials, via a combination of PCTs, projections, and isospectral extensions. 

\section{Conclusions}
In this manuscript, we have shown via an explicit construction that the Morse superpotential can be isospectrally deformed via the extension mechanism into the Scarf hypergeometric superpotential. As a result, we have demonstrated that there exists a path from a type-II to a type-I superpotential; thus, all known additive shape-invariant superpotentials are inter-related through a combination of PCTs, projections, and isospectral extensions.

\end{document}